# Adequacy of 5.9 GHz Band for Safety-Critical Operations with DSRC

Seungmo Kim[1], Min Jae Suh[2]
[1]Georgia Southern University, [2]Sam Houston University

## I. INTRODUCTION

Vehicle-to-Everything (V2X) communications has the potential to significantly bring down the number of vehicle crashes, thereby reducing the number of associated fatalities. Today, the two key radio access technologies (RATs) that enable V2X communications are Dedicated Short Range Communications (DSRC) and Cellular-V2X (C-V2X). DSRC is designed to primarily operate in the 5.9 GHz band (5.850-5.925 GHz). In December 2019, the US Federal Communications Commission (FCC) voted to allocate the lower 45 MHz (i.e., 5.850-5.895 GHz) out of the total 75 MHz for unlicensed operations to support high-throughput broadband applications (e.g., Wireless Fidelity, or Wi-Fi) [1]. While the reform is proposing to leave the upper 30 MHz (i.e., 5.895-5.925 GHz) for intelligent transportation system (ITS) operations (i.e., DSRC and C-V2X), it is also proposing to dedicate the upper 20 MHz of the chunk (i.e., 5.905-5.925 GHz) for C-V2X.

According to this plan, DSRC is only allowed to use 10 MHz of spectrum (i.e., 5.895-5.905 GHz). It has never been studied nor tested if 10 MHz would suffice for operation of the existing DSRC-based transportation safety infrastructure. As such, it has become urgent to understand how much impact of the FCC's 5.9 GHz band reform will be placed on the performance of existing ITS infrastructure.

## II. RELATED WORK

Current understanding on communications in the 5.9 GHz band is not sound enough to establish stable connections among infrastructure and vehicles.

*DSRC in High Traffic Density:* The performance of a DSRC broadcast system in a high-density vehicle environment was studied [2]; however, the assumption is too ideal to be realistic—i.e., the number of vehicles within a vehicle's communication range can be kept constant.

*Performance Evaluation Method:* A recent proposal combined a packet-level simulation model with data collected from an actual vehicular network [3]. Another recent work measured the performance of DSRC broadcast by using the geometrical methodology [9]. However, they did not study the impacts of internal and external bandwidth contention.

*Safety-Related Application:* A DSRC-based end of queue collision warning system was proposed [4]; however, it discusses a one-dimensional freeway model, which needs significant improvement for application to an intersection with two or more ways.

*External Bandwidth Contention:* The performance degradation of DSRC under interference from Wi-Fi was studied [5]; however, it lacks consideration of coexistence with C-V2X.

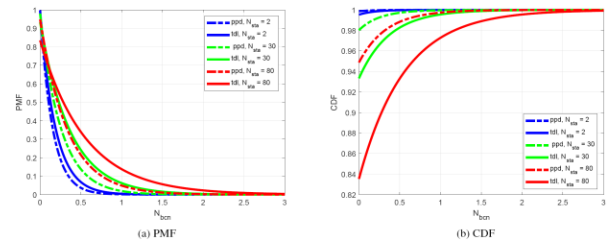

Fig. 1. Distribution of inter-reception time (IRT)

## III. CONTRIBUTIONS

Motivated from the scarcity in the literature, we propose a mechanism to "lighten" a DSRC network in order to fit in the reduced spectrum resiliently.

Fig. 1 describes the main result of our work. An inter-reception time (IRT) is defined as the time length between two consecutive safety messages. A longer IRT means a lower performance due to more safety message having been dropped. Fig. 1 shows that the proposed mechanism leads to a higher performance regardless of the network size (i.e., the number of nodes in a network).

Our contributions in our recent achievements are three-fold [6]-[15]: (1) We characterized the "decentralized" nature of a DSRC system by using the stochastic geometry; (2) As a standard prioritizing a message, we formulated the risk of an accident based on "distance to a danger source," which is a quantity that is easy to obtain by using the existing techniques and apparatus; (3) We established a generic analysis framework based on the stochastic geometry, in order to model a DSRC network's performance under external interference.